# Coherent coupling of molecular resonators with a micro-cavity mode

A. Shalabney[1], J. George[1], J. Hutchison[1], G. Pupillo[2], C. Genet[1]* and T.W. Ebbesen[1]*

[1]ISIS & icFRC, University of Strasbourg and CNRS (UMR 7006), 67000 Strasbourg, France
[2]IPCMS (UMR 7504) & ISIS (UMR 7006), University of Strasbourg and CNRS (UMR 7006), 67000 Strasbourg, France
Contact emails: genet@unistra.fr and ebbesen@unistra.fr

**The optical hybridization of the electronic states in strongly coupled molecule-cavity systems have revealed unique properties such as lasing, room temperature polariton condensation, and the modification of excited electronic landscapes involved in molecular isomerization. Here we show that molecular *vibrational* modes of the electronic ground state can also be coherently coupled with a micro-cavity mode at room temperature, given the low vibrational thermal occupation factors associated with molecular vibrations, and the collective coupling of a large ensemble of molecules immersed within the cavity mode volume. This enables the enhancement of the collective Rabi-exchange rate with respect to the single oscillator coupling strength. The possibility of inducing large shifts in the vibrational frequency of selected molecular bonds should have immediate consequences for chemistry.**

Intra-molecular vibrational motions can be described as a superposition of simple harmonic vibrations, so called molecular normal modes. For each of these modes, the atoms vibrate in specific directions that correspond to the observable vibrational transitions measured in infrared spectroscopy. The relatively high frequencies of molecular vibrational transitions $\omega_v$, fixed by the bond strength $f$ (typically of the order of $10^3$ N.m$^{-1}$) and the tiny atomic masses involved in the vibrations, immediately lead to two important features.

First, it is possible to perform *direct* resonant dipolar coupling by engineering micro-scaled cavities with a fundamental mode $\omega_c$ tuned to the molecular vibrational transitions. Then, as a consequence of their high frequencies in the infrared (IR) regime, vibrational resonances are characterized by small thermal occupation factors $n_\nu \sim e^{-\hbar\omega_\nu/k_B T} \sim 10^{-4}$, even at room temperature. This means that such molecular normal modes are in their ground state allowing coherent light-matter coupling in a straightforward manner.

In the following we demonstrate the coherent coupling between molecular vibrational transitions and an optical mode of a micro-cavity, leading to the possibility to swap, at room temperature, excitations coherently between the molecular oscillators and the optical mode. To do so, we have exploited two crucial features offered by polymers. First, the possibility to have an isolated, practically homogeneous, spectral signal associated to a specific vibrational molecular normal mode. Second, the capacity offered through the bulky extension of the polymer film inserted in a Fabry-Perot micro-cavity to have within a volume of strong optical confinement (i.e. the coherence volume of the cavity mode) a large number of resonators. The colocalization of the optical and mechanical modes induces a collective enhancement of the resonant coupling rate between the vibrational resonators and the cavity mode reaching the regime of strong coherent coupling. In other words, a macroscopic coherent mechanical mode is now generated by strong coupling.

## Results

### Hamiltonian description

Infrared spectra associated with gas-phase molecules usually display features where rotational transitions are coupled to vibrational ones. The resulting well-known complexity of rovibrational molecular spectra leads to spectral components separated by wave numbers less than 10 cm$^{-1}$. Still, there are specific environments where molecules can display much simpler spectra from which it is possible to select and manipulate chosen vibrational normal modes. Polymeric phases are in this context particularly interesting to explore since free rotations of molecular moieties are frozen-out and the excitation spectrum of the polymer is determined solely by electronic and vibrational contributions. One should also consider low frequency vibrations of the polymer lattice itself. Yet, given the small wave numbers for such lattice vibrations (less than ca. 100 cm$^{-1}$) compared to the vibrations of individual bonds (ca. 1000 cm$^{-1}$), the two classes of motions can be clearly separated[1]. In this regime therefore, vibrational spectra of polymers display normal mode transitions sufficiently isolated from the background to be tuned properly to a given cavity resonance, as presented on Fig. 1.

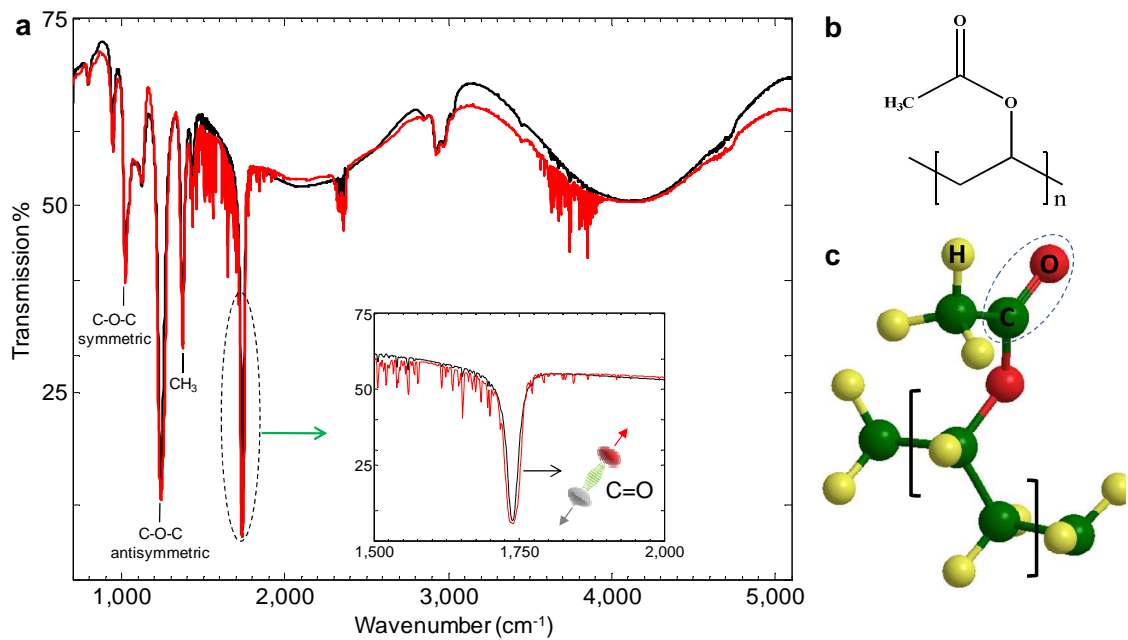

**Figure 1 | Polymer vibrational spectrum. a,** Transmission spectrum of polyvinyl acetate (PVAc) thin layer deposited on Ge substrate. The thickness of the film is about 2 microns and the measurement was performed at normal incidence. The measured transmission is normalized to free-space transmission. The black line fits the data modeling the polymer dispersion by ideal damped harmonic oscillators (see Supplementary Note 2). The inset shows the absorption band of the PVAc due to the (C=O) bond stretching band around 1740 cm$^{-1}$ with the same fit (black line). **b,** Chemical structure of a single PVAc monomer unit. **c,** Three dimensional structure of one PVAc monomer showing the (C=O) bond.

This situation allows us to define the molecular Hamiltonian from a "double adiabatic approximation" where a slow component (inter-molecular lattice vibrations) is separated from a fast subsystem describing intra-molecular motions[2]. In our situation, the fast component of the whole Hamiltonian is the only relevant one. We can then perform a second adiabatic (Born-Oppenheimer) approximation in order to separate the vibrational and electronic degrees of freedom. This separation leads to the definition of the vibrational dipole operator corresponding to the dependence on nuclear coordinates $Q$ of the expectation value of the dipole moment $\langle\hat{p}\rangle(Q)$ in the electronic state considered.

We emphasize that our coupling scheme only involves the fundamental electronic state. The low vibrational occupancy number implies moreover that only the fundamental level of the vibrational spectrum is populated. At such low excitations, the molecular vibrations occur within a mean electronic potential that is well treated in the harmonic approximation. We are thus finally dealing with a mechanical normal mode in its harmonic quantum ground state, which constitutes the engaging feature of molecular vibrations in this context - see Fig. 2(b).

We will further assume that the change in the ground state dipole moment when interacting with the cavity light mode can be limited to fundamental transitions, leaving aside higher order combination transitions and overtones. This corresponds to a simple first-order expansion

$$\langle\hat{p}\rangle(Q) = \langle\hat{p}\rangle_0 + \left(\frac{\partial\langle\hat{p}\rangle}{\partial Q}\right)_0 \cdot \hat{Q} \qquad (1)$$

of the vibrational dipole operator with respect to the equilibrium nuclear configuration in the harmonic mean potential of the electronic ground state (indicated by the subscript 0)[3]. The first term corresponds to the static dipole moment of the molecule at this equilibrium nuclear position, which does not contribute to the transition. The second term is involved in vibrational transitions induced by the resonant field, following the simple vibrational selection rule $\Delta v = \pm 1$.

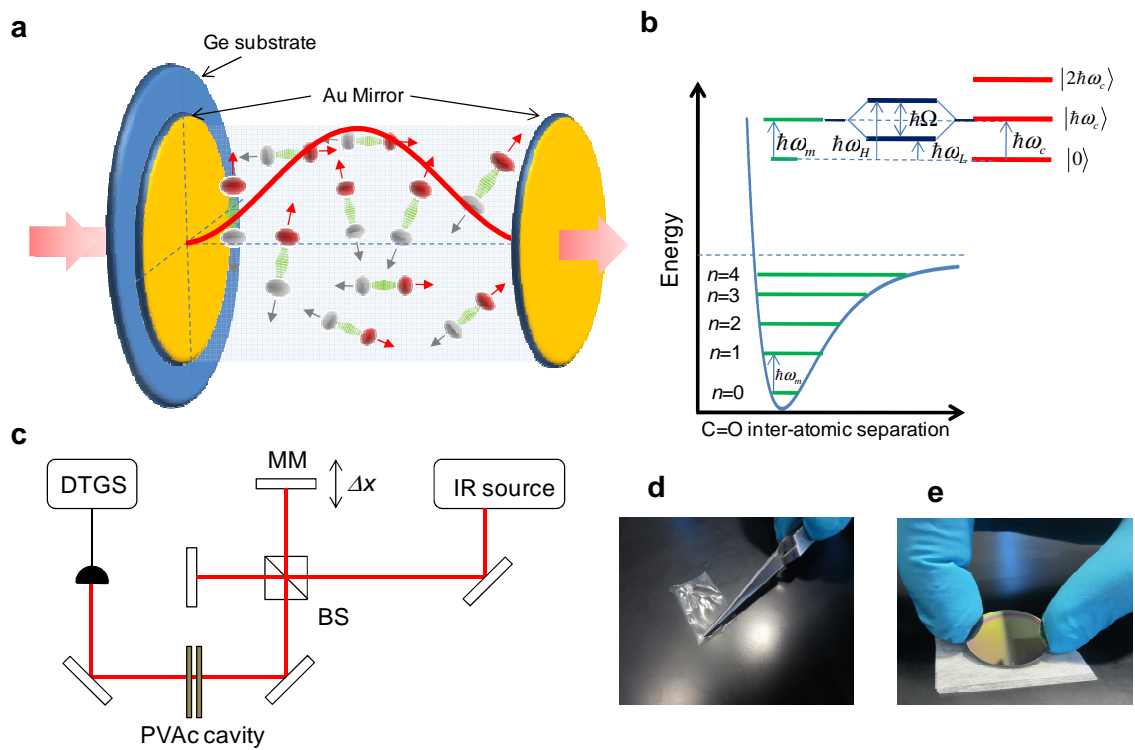

**Figure 2 | Microcavity and Experimental setup. a,** Schematic illustration of the microcavity used to strongly couple the (C=O) vibrational band to IR radiation. A thin (~ 2 μm) layer of PVAc is sandwiched between two symmetrical thin (10 nm ) Au mirrors deposited on a Ge substrate (see supplementary Note 1). The (C=O) bonds are depicted as mechanical oscillators inside the cavity with arbitrary orientations. The red thick curve describes the electric field intensity spatial distribution for the first cavity mode tuned in resonance with the (C=O) vibrational transition. **b,** Vibrational energy diagram in the anharmonic potential of the fundamental electronic state. The inset on the right top shows the coupling scheme between the fundamental vibrational mode and the first optical mode $\hbar\omega_c$ of the cavity and therefore forming two polariton branches ($\hbar\omega_H$ and $\hbar\omega_L$) separated by the Rabi energy. **c,** The experimental setup is based on a single-beam FT-IR system that records on a DTGS (Deuterated Triglycine Sulfate ) detector an interferogram generated from a movable mirror (MM). The interferogram is Fourier transformed to provide the actual vibrational spectrum. **d,** Photographic image of a free-standing PVAc layer, clearly showing the continuous character of the film. **e,** Photographic image of the cavity used in the experiments.

Under these assumptions, it is straightforward to treat in the dipole approximation the conservative interaction between a single molecular vibrational (one-dimensional) mode and the field in the cavity. This leads to the following Hamiltonian describing the coherent coupling regime:

$$H = \tfrac{1}{4}\hbar\omega_v\{Q_v{}^2 + P_v{}^2\} + \tfrac{1}{4}\hbar\omega_c\{Q_c{}^2 + P_c{}^2\} - \hbar\Omega\, Q_v\, Q_c, \qquad (2)$$

where $Q_c = (a + a^\dagger)$ corresponds to the optical position quadrature related to the photon $a(a^\dagger)$ annihilation (creation) operators associated with the cavity field at the position **r** of the molecular vibration bond, and $Q_v = (b + b^\dagger)$ corresponds to the vibrational position quadrature related to phonon $b(b^\dagger)$ annihilation (creation) operators, as a consequence of the mean potential harmonic approximation. While Eq. (2) is typical for cQED-physics, we emphasize that here the $Q_v$ quadrature corresponds not to electronic oscillations, but to a (optically induced) motion of atoms within the molecules.

In this approximation, the coupling strength reads

$$\hbar\Omega = \gamma \left(\frac{\partial \langle \hat{p}\rangle}{\partial Q}\right)_0 \sqrt{\frac{\hbar\omega_c}{2\varepsilon_0 V}}\, Q_{\text{zpf}}, \qquad (3)$$

where $V$ is the cavity mode volume and $Q_{\text{zpf}} = \sqrt{\hbar/(2\mu\omega_v)}$ the zero-point fluctuation amplitude of the molecular oscillator, determined from the reduced mass $\mu$ involved in the bond vibration. We also account for an orientation factor $\gamma = \hat{\boldsymbol{\epsilon}}_v \cdot \hat{\boldsymbol{\epsilon}}_c$ between the field polarization $\hat{\boldsymbol{\epsilon}}_c$ and the transition dipole polarization $\hat{\boldsymbol{\epsilon}}_v$. Spectrally, the regime of coherent coupling is seen as an avoided crossing between the two coupled modes at resonance. This splitting corresponds to the definition of two new normal modes for the coupled system with frequencies shifted from the individual (uncoupled) modes and separated by a vacuum Rabi energy equal to the coupling strength at resonance. This picture however neglects mechanical damping and cavity decay. In reality, the condition to reach such a strong coupling regime for a single oscillator is particularly stringent since $\hbar\Omega$ is well below 1 µeV for typical molecular vibrational transitions in the vacuum field of a micro-cavity. Therefore, this cannot exceed radiative and non-radiative damping rates of the system, including cavity losses. As well known however, the situation becomes different when several resonators are coupled to the same cavity mode.

Indeed, the spatially coherent single mode cavity field drives all the coupled resonators in phase with each other. This induces coherence among the resonators within the whole mode volume and leads to the definition of a macroscopic collective dipole, made of a large number of coupled transitions over which the excitation is delocalized. In the simplest case, neglecting all sources of inhomogeneous broadening[4], the strong coupling dynamics is defined within a two-level subspace made of collective ground and first excited states through an enhanced coupling rate $\hbar\Omega\sqrt{N}$, where $N$ is the ensemble of indiscernible resonators. Considering the very large number of resonators available in condensed systems, this enhancement can rise orders of magnitude, as observed in particular in organic systems[5-15]. It thus leads to conditions much easier to fulfill in order to enter and exploit the regime of coherent strong coupling[16].

**Experimental demonstration**

To test these ideas and demonstrate their consequences, we choose a polymer with a vibration band that is well isolated from other modes. Polyvinyl acetate (PVAc) has such a feature where the (C=O) bond has a symmetric stretching frequency at 1740 cm$^{-1}$ (215 meV) as shown in Fig. 1. The other peaks are due to other fundamental vibrational modes that are well-assigned, as illustrated in the caption[17]. The spectral line of the (C=O) vibration is remarkably close to a Lorentzian line shape (see inset in Fig. 1 a) revealing essentially a homogeneous intrinsic vibrational damping of $\hbar\Gamma_\nu \sim 3.2$ meV extracted from the (FWHM). This results in an associated mechanical quality factor $Q_\nu \sim 70$. Considering a typical 1 Debye dipole moment associated with the (C=O) bond in a polymeric phase[18] and a vacuum field amplitude $\sqrt{\hbar\omega/2\varepsilon_0 V}$ of ca. $6.3 \times 10^3$ V.m$^{-1}$ at the resonant frequency $\omega_c$, a conservative estimate of the coupling rate turns out to be of the order of $\hbar\Omega \sim 0.1$ μeV estimated from a diffraction limited $V \sim (\lambda/n)^3$ mode volume (*n* being the polymer background refractive index). This is much smaller than both the mechanical damping rate $\hbar\Gamma_\nu$ and the cavity decay rate $\hbar\kappa \sim 17$ meV (see below). In such conditions, it is thus impossible to strongly couple a single molecular vibration to the cavity vacuum field. However the extremely high density $\rho$ of (C=O) bonds in the polymer can in principle enable the formation of a collective dipole within the cavity mode volume, as discussed above. With one (C=O) bond per monomer, this density corresponds to approximately $\rho = 10^{21}$ cm$^{-3}$ (see Supplementary Note 4). This should indeed make reaching the strong coupling regime possible.

To demonstrate this, a specific Fabry-Perot cavity was engineered to have a first mode in the mid infrared (MIR) range resonant with the (C=O) vibrational transition (Fig. 1 a). This tuning requires a careful choice of substrates and metals forming the mirrors but eventually allowed us to demonstrate direct dipolar coupling between the cavity field and the molecular motion. The whole experimental setup, together with the description of the best material compromise, is shown in Fig. 2. Fourier-Transform Infra-red (FTIR) spectroscopy gives direct access to the spectral density of the transition through a phase-modulated signal transmitted through the cavity+polymer ensemble (see Supplementary Note 1). This interrogation mode allows the recording of angle-resolved spectral coherent responses of the coupled system.

As shown in Fig. 3, a Rabi anti-crossing is demonstrated at normal incidence in the dispersion relation of the cavity. The associated vacuum Rabi splitting $\hbar\Omega_R \sim 20.7$ meV exceeds all decoherence rates (evaluated above) and therefore corresponds to the regime of strong coupling. This leads to the formation of two new opto-vibrational modes for the system, as illustrated in Fig. 4. These new modes are the lower and upper polaritonic states and correspond to molecular vibrations dressed by the cavity vacuum field. From the widths of the associated spectral peaks, it is possible to give an estimate of the dephasing times of the dressed states which are 0.23 ps and 0.44 ps for the upper (UP) and lower (LP) polariton states, respectively. The generation of new hybrid vibrational states was further confirmed through the modification of the cavity fundamental modes as shown in Fig. 4. The obvious splitting in the field's distribution due to the strong coupling immediately indicates that the integrated absorption of the coupled system shows the same splitting as well, which is the unambiguous signature of the strong coupling regime (see Supplementary Figure S1). In our experiments, the observed splittings are probed at very low power and do not depend on it. This rules out any multi-photonic effects occurring in the experiments and reveals that the probe does not induce any AC-Stark effect in the system[19]. It thus confirms that the observed energy splittings are due to vacuum Rabi splitting only.

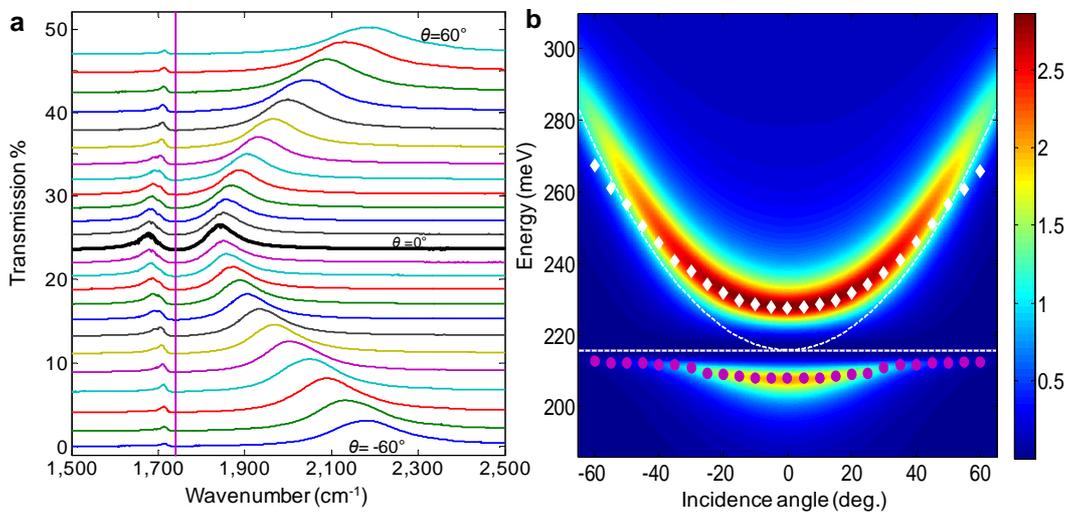

**Figure 3 | Cavity angular dispersion and strong coupling. a,** Cascade plot of measured transmission spectra through the Au-PVAc cavity as a function of the IR beam incidence angle. The spectra are vertically shifted every 5 degrees and the angular range covers -60 ; +60 degrees relatively to the cavity normal. At normal incidence ($\vartheta$=0°), the avoided crossing is clearly revealed as the signature of the strong coupling regime between the cavity mode and the (C=O) stretching mode (which position in an uncoupled situation is indicated by the vertical line). **b,** Color-plot of the cavity (Au-PVAc) dispersion calculated with parameters retrieved from the best transmission data fit at normal incidence (see supplementary Note 2). White diamonds and purple circles correspond respectively to the measured positions of the upper (UP) and lower (LP) polaritons extracted from the data displayed in **a**. Dashed curve and dashed horizontal line show respectively the dispersion of the empty cavity and (C=O) vibrational mode (see supplementary Note 2). The dispersion of the empty cavity was calculated by deactivating vibrational contributions and considering the background refractive index of the polymer. The crossing point between the dashed curves at normal incidence corresponds to the careful tuning between the first mode of the cavity with the (C=O) bond stretching mode. The Rabi splitting at the crossing point at normal incidence reaches 20 meV.

The relatively high value for the Rabi splitting is the clear signature of the collective coupling with $\hbar\Omega_R = \hbar\Omega\sqrt{N}$, from which it is tempting to evaluate an *effective* number of resonators coupled within the cavity mode volume. This is most easily given as an *effective* concentration $\rho_C$ of coupled molecules. With the same figures given above, this concentration can be estimated to be of the order of $\rho_C \sim 4.4 \times 10^{20}$ cm$^{-3}$. This number turns out to be slightly smaller than the expected (C=O) bond density $\rho$ (see Supplementary Note 4). This discrepancy immediately points to the central fact that in the real situation, the actual distribution of the (C=O) bonds within the cavity must be accounted for, including the spatial overlap between the molecules and the cavity mode and the orientational distribution of the molecular dipoles with respect to the cavity field. In such a "non-symmetrical" coupling, the initial rate $\Omega$ must be replaced by an averaged one, necessarily leading to a reduced Rabi splitting $\hbar\langle\Omega_R\rangle < \hbar\Omega\sqrt{N}$. As discussed in[20], this can be understood as an effective coherence volume associated with a single molecular resonator smaller than the cavity mode volume.

**Discussion**

Exploiting specific properties of polymers, we have been able to demonstrate for the first time the regime of vibrational strong coupling. The microscopic nature of our individual resonators leads to a practically perfect thermal decoupling of the molecular vibrations. Indeed, with $Q_\nu \cdot \omega_\nu / 2\pi \gg k_B T / \hbar$, thermal decoherence can be neglected over more than one vibrational period. Thus, essentially due to the extremely tiny effective mass involved in the mechanical stretching mode of the (C=O) bond, coherent coupling is achieved at room temperature. In addition, the high mechanical product $Q_\nu \cdot \omega_\nu / 2\pi$ holds promises in the context of transient spectroscopy. Indeed, while our discussion was here limited to the electronic ground state and first vibrational transition, one can envision to actually pump transiently the vibrational manifold. This might lead to inverted population dynamics in connection with polariton vibrational lasing[21].

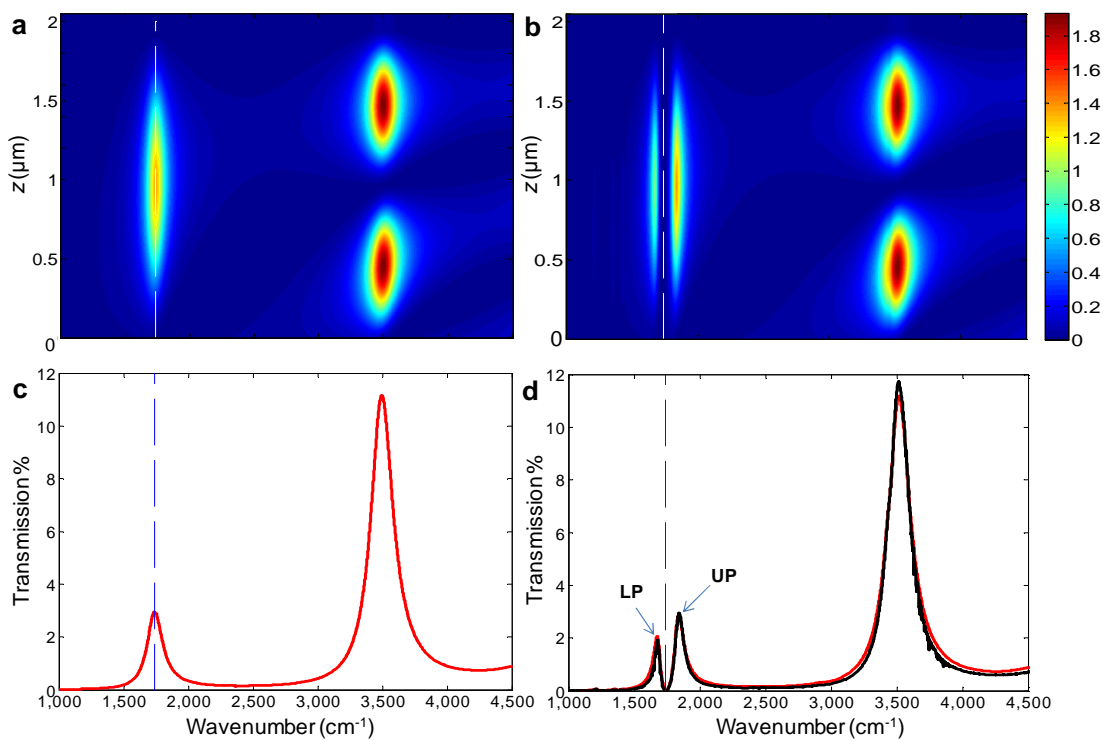

**Figure 4 | Strong coupling and intra-cavity field distributions. a,** Color-plot of the evolution of the intensity distribution inside the cavity in wavenumber. The vertical axis (*z*) scaled in mm is perpendicular to the cavity plane, with the first Au mirror at *z*=0. The thicknesses of both Au mirrors are 10 nm and the PVAc layer thickness is 1930 nm, values that were retrieved from the best fits. The intensity distribution is calculated in the situation of an *uncoupled cavity* where vibrational transitions within the polymer are deactivated, leaving only the non-dispersive background response of the polymer (see supplementary Note 2). The cavity polarizability is assumed to be homogenous and isotropic and the incidence angle is taken equal to zero. Vertical dashed line corresponds to the (C=O) vibration. **b,** Similar evaluation this time for the *strongly coupled* cavity where all the vibrational bands of PVAc are considered. The redistribution of the field into two new normal modes inside the cavity is clearly seen in the vicinity of the (C=O) vibrational band. In both cases, the second cavity mode is seen at higher wavenumber (ca. 3500 cm$^{-1}$) and characterized by two maxima across the cavity ($\lambda$-mode). The large differences between the first and second mode intensities are due to the mirrors dispersion. **c,** Transmission spectrum of the uncoupled cavity at normal incidence. **d,** Transmission spectrum of the coupled cavity at normal incidence (solid black curve) and associated theoretical fit (red curve). Here, the PVAc polarizability was retrieved from the measured transmission of the bare PVAc film (see Supplementary Note 2). Dashed vertical line indicates the (C=O) vibrational band. The signature of the strong coupling between the (C=O) band and the first cavity mode is clearly seen in such static transmission spectra by the new normal modes. All fit procedures and field calculations are detailed in the supplementary Notes 2 and 3 respectively.

Finally, because it involves dressed collective modes through the colocalization of the cavity field and the vibrational modes, large coupling rates with ratios $\Omega_R / \kappa$ close to 1 can be reached. This could lead to non-linear behavior in the infrared regime similar to that recently demonstrated for polariton Bose-Einstein condensation[22,23] in the optical regime.

The strong coupling of vibrational modes demonstrated here could have profound consequences for chemistry, as well as biochemistry. We have already shown that the rate and yield of a chemical reaction can be modified by strongly coupling an electronic excited state to the vacuum field[6]. In that case, the reaction involved a light induced isomerisation, a structural transformation of individual photochromic molecules, electronically strongly coupled in the optical regime. However most chemistry is done in the ground state and starts by bond breaking and formation. Therefore the modification of bond strengths in the ground state by strong coupling to molecular vibrations could open many possibilities in chemical reactivity, catalysis and site selective reactions. For instance, the optical resonance could be selectively tuned to the vibration of a bond targeted for dissociation. A reduction of the vibrational frequency through hybridization will most likely imply a weakening of the bond strength $f$ since $\omega \propto \sqrt{f/\mu}$. The ground state energy landscape governing the chemistry may be significantly modified. As an example of important chemical functional groups, the carbonyls (C=O), coupled in this study, play a central role in amide bonding in peptides and as coordinating units in metalloenzymes, as ligands in organometallic and coordination complexes, and as the active site in many industrial and pharmaceutical syntheses. For instance the reaction between benzaldehyde with phenylhydrazine to give a hydrazone, shown in Fig. 5, involves the breaking of the (C=O) bond and therefore its rate and possibly yield could be modified by such bond weakening through strong coupling. Of course this approach is not limited to the carbonyl stretch, any IR active mode of a molecular functional group could be coupled to a light mode in the way shown here.

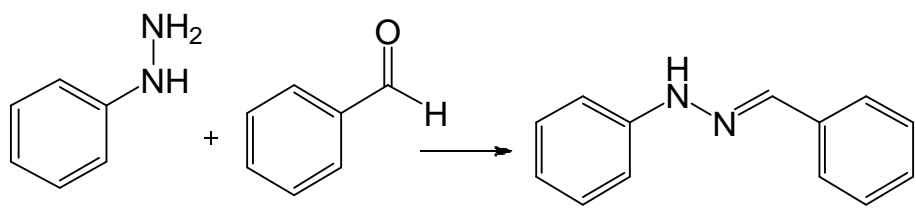

**Figure 5 | chemical reaction involves C=O bond breaking.** Benzaldehyde reacts with Phenylhydrazine to give a Hydrazone demonstrating a chemical reaction in which C=O bond breaking is involved.

The possibility of modifying chemical reaction rates, as described above, seems plausible based on several earlier experiments where bulk properties were modified by strong coupling, such as the already cited photochemical reaction, work-function and the ground state energy [6,24,25]. However the actual mechanism of how the strong coupling modifies the molecular material properties is still not clear and it might be counter-intuitive that the collective coupling induced in such systems will affect the properties of individual molecules. Further theory is indeed needed on such topic that can handle the complexity of strongly coupled molecular systems in order to be able to fully understand the potential of light-matter strong coupling for molecular science.

# References

1. Bower, D. I. & Maddams, W. F. *Vibrational Spectroscopy of Polymers.* (Cambridge University Press, Cambridge, 1992).

2. Voltz , R. Theory of molecular decay processes. *Org. Mol. Photophysics* **2,** 217-302 (1975).

3. Barron, L. *Molecular Light Scattering and Optical Activity.* (2nd Ed. Cambridge University Press, Cambridge, 2004).

4. Houdré, R., Stanley, R. P. & Ilegems, M. Vacuum-field Rabi splitting in the presence of inhomogeneous broadening: Resolution of a homogeneous linewidth in an inhomogeneously broadened system. *Phys. Rev. A* **53,** 2711-2715 (1996).

5. Agranovich, V. M., Gartstein, Y. N. & Litinskaya, M. Hybrid Resonant Organic_Inorganic Nanostructures for Optoelectronic Applications. *Chem. Rev.* **111,** 5179-5214 (2011).

6. Hutchison, J. A., Schwartz, T., Genet, C., Devaux, E., Ebbesen, T. W. Modifying Chemical Landscapes by Coupling to Vacuum Fields. *Angew. Chem.* **51,** 1592-1596 (2012).

7. Pockrand, I., Brillante, A. & Möbius, D. Exciton–Surface Plasmon Coupling: An Experimental Investigation. *J. Phys. Chem.* **77**, 6289-6295 (1982).

8. Lidzey, D. G. *et al.* Strong exciton–photon coupling in an organic semiconductor microcavity. *Nature* **395,** 53-55 (1998).

9. Schwartz, T., Hutchison, J. A., Genet, C., Ebbesen, T. W. Reversible switching of ultrastrong light-molecule coupling. *Phys. Rev. Lett.* **106,** 196405-196408 (2011).

10. Aberra Guebrou, S. *et al.* Coherent emission from a disordered organic semiconductor induced by strong coupling to surface plasmons. *Phys. Rev. Lett.* **108**, 066401 (2012).

11. Wang, S. *et al.* Quantum yield of polariton emission from hybrid light-matter states. *J. Phys. Chem. Letters* **5**, 1433-1439 (2014).

12. Wang, S. *et al.* Phase transition of a perovskite strongly coupled to the vacuum field. *Nanoscale* **6**, 7243-7248 (2014).

13. Berrier, A. *et al.* Active control of the strong coupling regime between porphyrin excitons and surface plasmon polaritons. *ACS Nano* **5**, 6226-6232 (2010).

14. Hakala, T. K. *et al.* Vacuum Rabi splitting and strong-coupling dynamics for surface plasmon polaritons and rhodamine 6G molecules. *Phys. Rev. Lett.* **103**, 053602 (2009)

15. Vasa, P. *et al.* Real-time observation of ultrafast Rabi oscillations between excitons and plasmons in metal nanostructures with J-aggregates. *Nature Photon.* **7,** 128-132 (2013).

16. Kaluzny, Y., Goy, P., Gross, M., Raimond, J. M. & Haroche, S. Observation of Self-Induced Rabi Oscillations in Two-Level Atoms Excited Inside a Resonant Cavity: The Ringing Regime of Superradiance. *Phys. Rev. Lett.* **51,** 1175-1178 (1983).

17. Terui, Y. & Hirokawa, K. Fourier transform infrared emission spectra of poly(vinyl acetate) enhanced by the island structure of gold. *Vib. Spec.* **6,** 309-314 (1994).

18. Koenig, J. *Spectroscopy of Polymers*. (2nd ed. Elsevier Science Inc., New-York, 1999).

19. Schwartz, T. *et al.* Polariton Dynamics under Strong Light-Molecule Coupling. *Chem. Phys. Chem.* **14,** 125-131 (2013).

20. Haroche, S. Fundamental Systems in Quantum Optics. Proceedings of the Les Houches Summer School. (Les Houches, session LIII, Elsevier Science Publishers, Amsterdam, 1992).

21. Kéna-Cohen, S. and Forrest S. R. Room-temperature polariton lasing in an organic single-crystal microcavity. *Nature Photon.* **4**, 371-375 (2010).

22. Plumhof, J. D., Stöferle, T., Mai, L., Scherf, U., & Mahrt, R. F. Room-temperature Bose-Einstein condensation of cavity exciton-polaritons in a polymer. *Nature Mat.* **13**, 247-252 (2014).

23. Daskalakis, K. S., Maier, S. A., Murray, R., & Kéna-Cohen, S. Nonlinear interactions in an organic polariton condensate. *Nature Mat.* **13**, 271-278 (2014).

24. Hutchison, J. A., *et al.*, Tuning the Work-Function Via Strong Coupling, *Adv. Mat.* **25**, 2481-2485 (2013).

25. Canaguier-Durand, A. *et al.* Thermodynamics of Molecules Strongly Coupled to the Vacuum Field. *Angew. Chem. Int. Ed.* **125,** 10727-10730 (2013).

**Acknowledgments**

The authors thank C. Genes, P.S. Julienne and J. Moran for discussions and acknowledge support from the ERC (Grants 227557 and 307688), the International Center for Frontier Research in Chemistry (icFRC, Strasbourg), the ANR Equipex "Union" (ANR-10-EQPX-52-01) and the Labex NIE projects (ANR-11-LABX-0058_NIE) within the Investissement d'Avenir program ANR-10-IDEX-0002-02, and EOARD.

**Authors contribution**

All authors contributed to all aspects of this work.

# Supplementary information

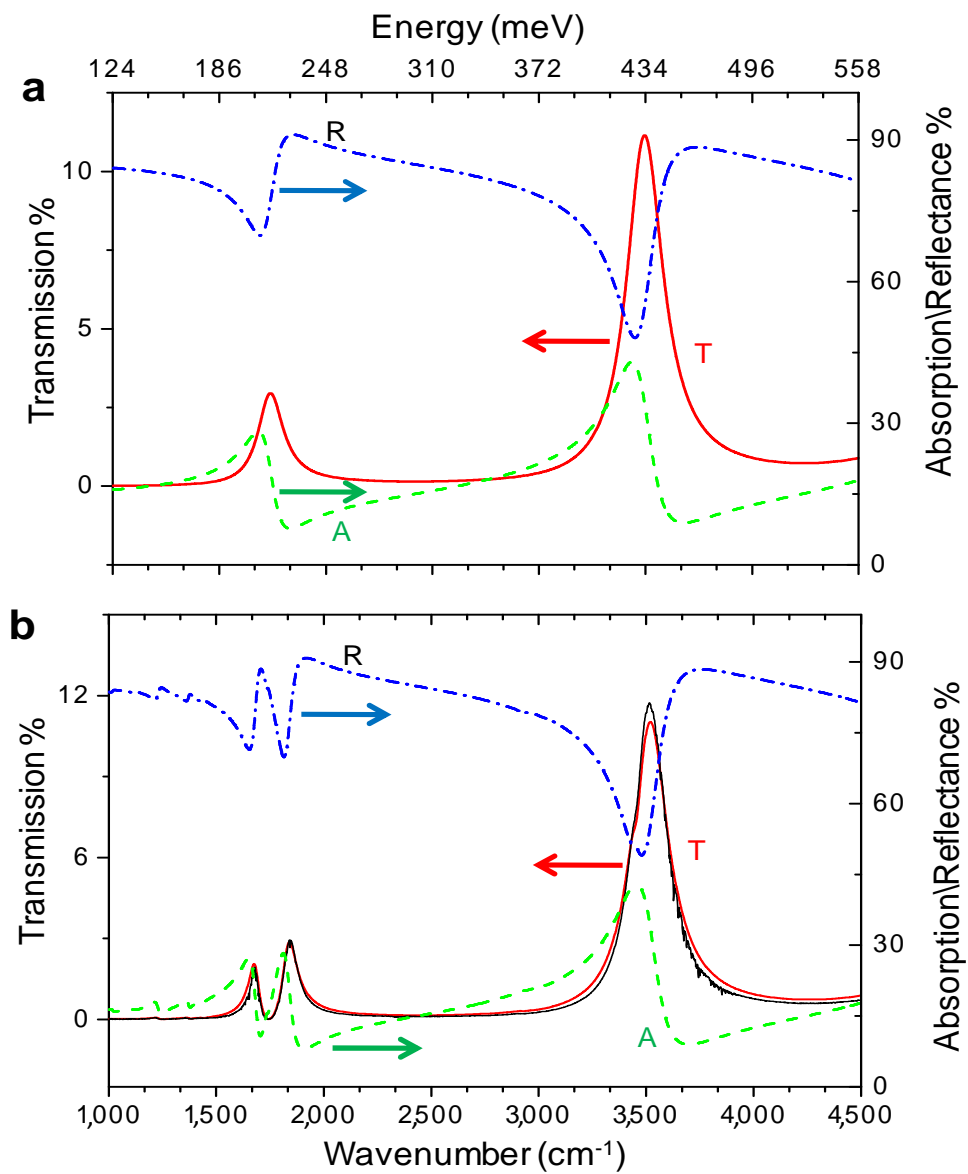

**Supplementary Figure S1| Transmission, absorption, and reflection of the un-coupled and coupled cavity.** **a**, Transmission (red solid line), absorption (green dashed line), and reflection (blue dashed dotted line) of the uncoupled cavity using the best fit parameters from the measured transmission spectrum. **b**, Transmission (red solid line) with the measured spectrum (black solid line), absorption (green dashed line), and reflection (blue dashed dotted line) of the coupled cavity.

# Supplementary Note 1

**Cavity preparation and dispersion measurements**

An approximately 10 nm- thick Au mirror was sputtered on a clean Ge substrate. Then, a polymer (Poly Vinyl Acetate or PVAc) film was deposited by spin-casting (4230 rpm) to form a layer of about 2 microns thickness tuned to overlap the first cavity mode with the (C=O) bond stretching band of PVAc at 1740 cm$^{-1}$.

PVAc (MW: 140000) was dissolved in Toluene (15 wt. %) and mechanically steered at 100 °C for 40 hours, cooled to room temperature and passed thru a 0.22 µm Nylon filter prior to spin-casting. Finally, the cavity was formed by sputtering a second Au layer (10 nm) directly on top of the polymer. The thickness of the Au mirrors was optimized to obtain sufficient intensity in the transmission spectra in the required spectral region. Due to the significant increase of both the real and imaginary parts of the dielectric constant of Au in the infra-red (IR) region, a compromise was necessary between the cavity quality factor and the transmission mode intensity.

To measure the net polymer absorption without the cavity, another sample of the same thickness was prepared by spin-casting the PVAc solution directly on top of a clean Ge substrate.

The spectra of the cavity were acquired by standard FTIR (Fourier transform infra-red) spectrometer (Nicolet 6700) in transmission mode (description of the measurement set-up is given in Fig. 2c in the main text). Prior to every measurement, a background was acquired in order to normalize the actual measurement and avoid baseline instability. All the measurements were performed with a resolution of 1 cm$^{-1}$ and averaged over 128 scans to enhance the signal to noise (SNR) in the spectral range 400-7400 cm$^{-1}$.

The dispersion of the two hybrid states of the coupled system was measured by varying the incidence angle in the range from -60° to +60° (as shown in Fig. 3 in the main text). The position of the bare cavity mode was tuned by varying the in-plane wave vector $k_x$ with the incidence angle ( $k = 10^4 /(2nd \cos \theta_i)$ ) whereas k is the first cavity mode wave-vector in cm$^{-1}$, n the background polymer refractive index, d the cavity thickness, and $\theta_i$ the incidence angle.

## Supplementary Note 2

**Theoretical fit of the transmission spectra**

In order to simulate the optical response of the polymer in the IR region, the Lorenz model was used to describe the molecular polarizability. Since the absorption of the uncoupled polymer was obtained from the transmission of a bare polymer thin film on Ge substrate, an interference pattern was observed with values higher than 100% (negative absorbance values). This interference effect is observed when the measurement is normalized to the substrate transmission. In order to estimate the real absorption of the bare film, the Lorenz model with multiple resonances in the relevant spectral range was assumed as follows[1]:

$$\varepsilon(k) = \varepsilon_B - \sum_{j=1}^{N} \frac{f_j}{k^2 - k_{0j}^2 + ik\Gamma_j} \quad (1)$$

Here, $\varepsilon_B$, $f_j$, $k_{0j}$, and $\Gamma_j$ are respectively the background dielectric contribution, oscillator strength, resonance wave vector, and the phenomenological damping constant of vibrational band $j$. The absorption intensity of the band is determined by both $f_j$ and $\Gamma_j$ whereas the FWHM is solely governed by $\Gamma_j$. In our fit procedure, all these parameters together with the thin film thickness were varied to obtain the best fit with the experimental measurements. The best fit parameters for the (C=O) bond absorption band were $f_j = 50 \times 10^3$, $k_{0j} = 1739 cm^{-1}$, $\Gamma_j = 13 cm^{-1}$ and the background refractive index (RI) is $n_B = \sqrt{\varepsilon_B} = 1.41$.

The dispersion of the Au mirrors, on the other hand, was modeled by the Lorenz-Drude equation (2). In this model, the contributions of the intra-band and the inter-band transitions are explicitly separated as can be seen in Eq. (2)[2]:

$$\varepsilon_m = \varepsilon_{free-electrons} + \varepsilon_{bound-electrons}$$

$$\varepsilon_{free-electrons} = 1 - \frac{f_0 \omega_p^2}{\omega(\omega - i\Gamma_0)} \quad ; \quad \varepsilon_{bound-electrons} = \sum_{j=1}^{K} \frac{f_j \omega_p^2}{(\omega_j^2 - \omega^2) + i\omega\Gamma_j} \quad (2)$$

In the last two terms, $\omega_p$ is the plasma frequency, $K$ is the number of oscillators involved, each with frequency $\omega_j$, strength $f_j$ and life time $1/\Gamma_j$.

Since the thickness of the Au mirrors used in our study is very small, the dielectric constant becomes size dependent. For noble metals, this applies to structures smaller than the conduction electron mean-free path which is roughly 20 nm or less in the smallest dimensions[3]. The most significant influence of the size-dependence on the dielectric function is broadening of the plasmon width due to electron scattering at the boundaries. In order to describe this effect, it is convenient to consider the same Lorenz-Drude model as in Eq. (2) with the introduction of additional damping term $\Gamma_S$ to the normal damping $\Gamma_0$ which is proportional to $v_F/L$ where $v_F$ and $L$ are the Fermi velocity and the mirror thickness respectively. Within our theoretical procedure, this additional term was varied to improve the fit with the experimental measurements. The parameters of Au were adopted from Rakic *et al.* (see Table 1)[2]. The overall damping term $\Gamma_{tot} = \Gamma_0 + \Gamma_S$ which gave the best fit with the experimental results was found to be $2.5\Gamma_0$.

**Table 1** Gold dielectric constant parameters according to Eq. 2, all the parameters are given in eV units.

| $\omega_p$ | $f_0$ | $\Gamma_0^*$ | $f_1$ | $\Gamma_1$ | $\omega_1$ | $f_2$ | $\Gamma_2$ | $\omega_2$ | $f_3$ | $\Gamma_3$ | $\omega_3$ | $f_4$ | $\Gamma_4$ | $\omega_4$ | $f_5$ | $\Gamma_5$ | $\omega_5$ |
|---|---|---|---|---|---|---|---|---|---|---|---|---|---|---|---|---|---|
| 9.03 | 0.76 | 0.05 | 0.02 | 0.24 | 0.41 | 0.01 | 0.34 | 0.83 | 0.07 | 0.870 | 2.96 | 0.60 | 2.49 | 4.30 | 4.38 | 2.21 | 13.32 |

*The total damping term was considered as 0.125 in the simulations.

The transmission of the entire cavity was calculated using the standard 2x2 propagation matrices for a multilayer system embedded between two semi-infinite dielectric, isotropic, and homogenous media[4]. Correction for the transmission from the rear Ge/air interface was performed using the transmission spectrum of the bare Ge substrate.

After retrieving the optimal parameters of the cavity, the dispersion for the empty cavity was estimated by eliminating the absorption band's contributions from the dielectric constant. The splitting in reflection and absorption are compared to the transmission measurements as shown in Fig. S1. The observed values of transmission as well as the spectral shapes are in excellent agreement with the theoretical calculations. However, from the reflection and absorption spectra, one can see that the splitting is dependent on the measurement type as was confirmed in our previous works[5]. The observed splitting in transmission $(\Omega_T)$ is 167 cm$^{-1}$ (20.7 meV), whereas the splitting in absorption $(\Omega_A)$ and reflection $(\Omega_R)$ were slightly smaller (162 cm$^{-1}$ and 161 cm$^{-1}$ respectively). The ratio $(\Omega_T/\Omega_A)$ is very close to unity which is an indication of high coupling strength[5].

In the measured spectra, one can note some asymmetries between the LP and UP spectra at normal incidence (Fig. 3a in the main text). The small spectral asymmetry of the two hybrid states with respect to the vibrational band frequency is mainly caused by the background effect of the polymer. We assumed for the uncoupled cavity that the background polarizability is frequency independent. In fact, the non-uniform free spectral range (FSR), measured for higher modes of control cavities made on glass substrates, showed that the background slightly disperses through the spectral range. The intensities and shapes of the polariton profiles, on the other hand, are mainly affected by the dispersion of the metallic mirrors as well as the vibrational band profile. From the excellent fit of the measured spectra (Fig. 4d), we can ascertain the difference in polariton peak intensities to the dispersive response of the metal. Finally, added to the dispersion of the metal, the original asymmetry in the (C=O) bond vibrational mode explains the differences in the shapes of the two polariton spectral signatures.

## Supplementary Note 3

**Field distribution**

The electromagnetic field intensity calculations were based on a general algorithm considering the well known 4x4 propagation matrices in stratified media. In our case, the layers constituting the cavity, namely the two thin Au mirrors and the polymer layer were assumed to be isotropic and homogeneous and the incident field to be arbitrary polarized. In all the calculations, the experimental beam divergence and the approximately Gaussian beam shape of the FTIR were taken into account.

## Supplementary Note 4

**Estimation to the number of (C=O) bonds inside the cavity volume**

The (C=O) density in the cavity can be calculated by dividing the PVAc density (1.18 g/cm$^3$) by the PVAc monomer weight (86.09 g/mole) and multiplying by Avogadro's number, yielding 8.25x10$^{21}$ (C=O) per cm$^3$. This implies that in the mode volume of the cavity (ca. 10$^{-4}$ cm$^3$) there are about 8.25x10$^{17}$ (C=0) bonds.

## Supplementary References

1. Oughstun, K. E. & Cartwright, N. A. On the Lorentz-Lorenz formula and the Lorenz model of dielectric dispersion. *Opt. Express* **11,** 13, 1541-1546 (2003).

2. Rakic, A. D. , Djurišic, A. B. , Elazar, J. M. & Majewski, M. L. Optical properties of metallic films for vertical-cavity optoelectronic devices. *Appl. Opt.* **37,** 22, 5271-5283 (1998).

3. Kreibig, U. Electronic properties of small silver particles: the optical constants and their temperature dependence. *J. Phys. F: Metal Phys.* **4,** 999-1014 (1998).

3. Born, M. & Wolf, E. *Principles of Optics: Electromagnetic Theory of Propagation, Interference and Diffraction of Light.* (7th Ed., Cambridge press, 1999)

5. Schwartz, T., Hutchison, J. A., Genet, C. & Ebbesen, T. W. Reversible switching of ultrastrong light-molecule coupling. *Phys. Rev. Lett.* **106,** 196405-196408 (2011).